\documentclass[apj]{emulateapj}

\shorttitle{Gap Formation}
\shortauthors{Duffell \& MacFadyen}

\begin{document}

\title{Gap Opening by Extremely Low-Mass Planets in a Viscous Disk}

\author{Paul C. Duffell and Andrew I. MacFadyen}
\affil{Center for Cosmology and Particle Physics, New York University}
\email{pcd233@nyu.edu, macfadyen@nyu.edu}

\begin{abstract}

By numerically integrating the compressible Navier-Stokes equations in two dimensions, we calculate the criterion for gap formation by a very low mass (${q \sim 10^{-4}}$) protoplanet on a fixed orbit in a thin viscous disk.  In contrast with some previously proposed gap-opening criteria, we find that a planet can open a gap even if the Hill radius is smaller than the disk scale height.  Moreover, in the low-viscosity limit, we find no minimum mass necessary to open a gap for a planet held on a fixed orbit.  In particular, a Neptune-mass planet will open a gap in a minimum mass solar nebula with suitably low viscosity (${\alpha \lesssim 10^{-4}}$).  We find that the mass threshold scales as the square root of viscosity in the low mass regime.  This is because the gap width for critical planet masses in this regime is a fixed multiple of the scale height, not of the Hill radius of the planet.

\end{abstract}


\section{Introduction}
\label{sec:intro}

The question of whether a gap will form in a planet-disk system is of critical importance in understanding its evolution \citep{kn12}.  The amount of gas in the planet's immediate vicinity will strongly determine the migration rate through the disk \citep{w97} and potentially even halt migration \citep{wh89,ll09,yl10}.  Gap opening can also dramatically affect accretion onto the secondary \citep{lp_main}.  As such, planetary evolution models are sensitive to the gap opening criterion, and it is therefore of great interest to know this criterion precisely.

Several conditions have been proposed for gap opening.  \cite{lp_main} proposed a stability limit, which is essentially a limit on the shear produced by steep pressure gradients in the disk.  The Rayliegh Stability criterion reduces to a condition that the planet's Hill radius be larger than the disk scale height:
\begin{equation}
R_H = r (q/3)^{1/3} > h,
\label{eqn:stab}
\end{equation}
where r is the orbital radius, $q = M_p/M_*$ is the mass ratio, and h is the disk scale height.  We will refer to this limit as the ``Strong Shock" limit.  This is because it is related to the requirement that a strong shock forms within a scale height of the planet's orbit.  In terms of a limit on the planet mass, (\ref{eqn:stab}) reduces to
\begin{equation}
M_p > 3 M_{Sh},
\label{eqn:hillradius}
\end{equation}
\begin{equation}
M_{Sh} = h c^2 / G = M_* / \mathcal{M}^3
\end{equation}
Here, $M_p$ is the mass of the secondary, $c$ is the local sound speed, $M_*$ is the primary mass, and $\mathcal{M} = r/h$ is the disk Mach number.  \cite{lp_main} also demonstrated a viscous limit, which requires that the gap be opened faster than viscosity can refill it:
\begin{equation}
q > { 40 \nu \over \Omega r^2 }
\end{equation}
where $q$ is the mass ratio, $\nu$ is the viscosity, $r$ is the distance between the primary and the secondary, and $\Omega$ is the orbital frequency.  In terms of the Shakura-Sunyaev viscosity $\nu = \alpha c h$ \citep{ss73}, this viscous limit is
\begin{equation}
M_p > M_{\alpha},
\end{equation}
\begin{equation}
M_{\alpha} = 40 \alpha \mathcal{M} M_{Sh}.
\label{eqn:oldvisc}
\end{equation}
Another important gap opening criterion is the inertial limit, which appropriate for migrating planets \citep{ll09,yl10}.  In this case, the gap opening rate must be faster than the secondary's migration rate.  \cite{wh89} calculated this limit:
\begin{equation}
M_p > M_I,
\end{equation}
\begin{equation}
M_I \sim {\Sigma h^2 \over \mathcal{M}} \sim M_{Sh} {\Sigma r^2 \over M_*}
\label{eqn:inertial}
\end{equation}
Here, $\Sigma$ is the surface density.  In typical protoplanetary disks, the disk mass ${\Sigma r^2}$ is much smaller than the primary mass, meaning that planets satisfying the strong shock criterion are essentially guaranteed to satisfy (\ref{eqn:inertial}).  Synthesizing these results, \cite{cm07} built an effective 1D description of the disk profile and used it to construct a gap opening criterion which combines the viscous limit with the strong shock limit:
\begin{equation}
1.1 \left( {M_p \over M_{Sh}} \right)^{-1/3} + 50 \alpha \mathcal{M} \left( {M_p \over M_{Sh}} \right)^{-1} < 1
\label{eqn:crida}
\end{equation}
Note that already the ``strong shock limit" has been relaxed to ${M_p > 1.3 M_{Sh}}$.  However, this picture cannot be accurate.  It has been demonstrated, both theoretically \citep{gr01,r02} and in direct numerical calculations \citep{drs2,disco} that planets with masses well below $M_{Sh}$ can still open gaps.

The point, as noted by \cite{r02b}, is that gap opening does require a dissipation mechanism such as a shock, but not necessarily a strong shock.  In fact, any perturbing mass on a fixed orbit in an inviscid disk will eventually open a gap.  However, in real situations with nonzero viscosity, ``eventually" may entail unrealistic timescales, i.e. longer than the viscous timescale or the migration timescale.

If the strong shock limit is not a necessary condition for gap opening, then migration effects become important, and the inertial limit (\ref{eqn:inertial}) would become relevant.  Additionally, three-dimensional effects could no longer be neglected, because the planet's radius of influence would not extend across the entire thickness of the disk.  Moreover, the viscous limit (\ref{eqn:oldvisc}) was derived assuming that the strong shock criterion was satisfied, so it might take a different form when applied to low mass secondaries.

To understand why this might be the case, we can estimate the rate of angular momentum transfer due to the planet and due to viscosity, as has been previously estimated by \cite{lp_main}:
\begin{equation}
\dot H_p \sim r c \Sigma h^2 \left({ h \over \Delta } \right)^3 \left({ M_p \over M_{Sh} } \right)^2 
\end{equation}
\begin{equation}
\dot H_{\alpha} \sim \nu \Sigma r^2 \sim \alpha c h \Sigma r^2,
\end{equation}
where $\Delta$ is the gap width, indicating that the tidal torque depends on how far the edge of the gap is from the secondary (This crude estimate can be improved upon by direct calculations of planet torque, see for example \cite{dl10}, \cite{r11}).  The gap opening criterion can be estimated by equating these torques, ${\dot H_p \sim \dot H_{\alpha}}$:
\begin{equation}
\left({ M_{gap} \over M_{Sh} } \right)^2 \sim \alpha \mathcal{M}  \left({ \Delta \over h } \right)^3
\end{equation}
If the gap width is of order the disk scale height ${\Delta \sim h}$,
\begin{equation}
M_{gap} \sim M_{Sh} \sqrt{ \alpha \mathcal{M} }
\label{eqn:scale_h}
\end{equation}
However, if the gap width is governed by the secondary's Hill radius, ${\Delta \sim R_H \sim r q^{1/3}}$,
\begin{equation}
M_{gap} \sim M_{Sh} ( \alpha \mathcal{M} ),
\label{eqn:scale_r}
\end{equation}
which is in agreement with the form found for large mass planets by \cite{lp_main}, our equation (\ref{eqn:oldvisc}).  This indicates that the scaling with $\alpha$ is dependent upon whether the width of the gap is governed by the scale height of the disk, or the Hill radius of the planet.  One might suspect that it is governed by whichever of the two is greater, meaning that if the secondary is small enough that its Hill radius is smaller than a scale height, the width of the gap is independent of the mass, and the scaling should correspond to (\ref{eqn:scale_h}).  This turns out to be correct, as we shall confirm in this work.

Using a direct numerical approach, we find the viscous limit for a low-mass secondary (one which does not satisfy the strong shock criterion).  We keep the planet at a fixed orbital radius and vary its mass and the disk viscosity, to empirically determine the gap opening criterion as a function of the dimensionless Shakura-Sunyaev viscosity parameter, $\alpha$.  Migration effects are intentionally neglected, so that we may isolate the dependence on viscosity alone.

The numerical integration is carried out using a moving computational mesh \citep{tess,disco}.  This reduces numerical viscosity and allows us to take long time-steps, thus making it possible to run the calculations for a large number of orbits, and hence to explore the low-viscosity regions of parameter space.

\section{Numerical Ingredients}
\label{sec:num}
The equations being solved are the compressible Navier-Stokes equations.  For an effective 2-D description of the gas, these equations are vertically integrated (e.g. $\Sigma = \int \rho dz$).  In other words, vertical structure is neglected, and vertically propagating modes are considered subdominant:

\begin{equation}
\partial_t \Sigma + \partial_i ( \Sigma v_i ) = 0
\end{equation}
\begin{equation}
\partial_t ( \Sigma v_j ) + \partial_i ( \Sigma v_i v_j ) + \partial_j P = F^{visc}_j + F^{grav}_j
\end{equation}
\begin{equation}
\partial_t E + \partial_i ( (E + P)v_i ) = (F^{visc}_j + F^{grav}_j) v_j
\end{equation}

Where $\Sigma$ is the surface density, $v$ is the velocity, P is the pressure, and $E$ is the energy density, ${E = 1/2 \rho v^2 + \epsilon_{int}}$.  The equation of state is taken to be adiabatic,
\begin{equation}
P = \epsilon_{int} ( \gamma - 1 )
\end{equation}
but we choose $\gamma = 1.001$ to make the equation of state nearly isothermal.

The source terms include the viscous force,
\begin{equation}
F^{visc}_j = \partial_i ( \rho ( \frac12 \nu (\partial_i v_j + \partial_j v_i) + \zeta \delta_{ij}\partial_k v_k ) ),
\end{equation}

where $\nu$ is the kinematic viscosity, and the gravitational force, $F_{grav}$, which is the force produced by both point masses.  These masses are each moved in a Keplerian circular orbit about a common center of mass.  In our implementation, the viscous force is re-expressed as a viscous flux and moved to the left hand side of the equation.

\subsection{DISCO Code}

DISCO is a finite-volume, moving-mesh hydrodynamics code which is specifically tailored to the study of gaseous disks.  DISCO utilizes a cylindrical grid which moves and shears azimuthally with the orbital velocity of the fluid.  It is based on the TESS code, which is a general moving-mesh method for solving hyperbolic systems \citep{tess,disco}.

The computational domain runs from $r = 0.4$ to $r = 1.6$ with 512 radial zones, which is about 25 zones per scale height in the vicinity of the planet's orbit.  The number of azimuthal zones varies with radius, in order to keep a fixed aspect ratio of 1:1.  This amounts to $N_{\phi} \approx 3200$ zones.

The calculation of the viscous terms is detailed in the appendix.  Viscosity on a moving mesh has already been implemented by \cite{mvisc} for a Voronoi tessellation, but because our time integration utilizes the method of lines, our implementation is much simpler.

Kinematic viscosity in the disk is assumed to be a constant, but we report our results in terms of the dimensionless parameter ${\alpha = \nu c_p h_p}$ (the subscript p indicates we are evaluating these quantities at the planet's orbital radius).  In other words, $\alpha$ is not spatially constant, but we report our results in terms of its value in the vicinity of the secondary.

\subsection{Background Flow} 

The initial conditions assume the Minimum Mass Solar Nebula (MMSN) \citep{h81}:
\begin{equation}
\Sigma(r) = \Sigma_0 (r_p/r)^{3/2}
\end{equation}
\begin{equation}
c(r) = c_0 (r_p/r)^{1/4}
\end{equation}
with orbital frequency set to balance pressure and gravitational forces:
\begin{equation}
\Omega^2(r) = \Omega_0^2(r) \left( 1 - 2 \sqrt{r/r_p}/\mathcal{M}^2 \right),
\end{equation}
\begin{equation}
\Omega_0(r) = \sqrt{ GM_*/r^3 }
\end{equation}
Because we keep the planet at a fixed orbit and the gas exerts no gravitational force, $\Sigma_0$ is arbitrary.  $c_0$ is fixed by the mach number:
\begin{equation}
c_0 = r_p \Omega_0 / \mathcal{M}.
\end{equation}

For the Mach number, we choose $\mathcal{M} = 20$, which corresponds to an orbital radius $r_p = 2.4$ AU.  The Mach number is not fixed as a function of orbital radius.

\subsection{Perturbing Potential}

The gravitational well of the planet is given by the simple Newtonian formula:
\begin{equation}
\Phi(s) = G M_p / s
\label{eqn:newtG}
\end{equation}
where $s$ is the distance from the planet to the fluid element.  However, we use a smoothed potential, for two reasons.  First, the formula (\ref{eqn:newtG}) is divergent, and its singularity is on the grid, which would make time integration impossible.  Secondly, in this effective 2-D treatment, (\ref{eqn:newtG}) would overestimate the secondary's influence within a scale height of the planet, because our two dimensional fluid elements actually feel a vertical average of this gravitational force.  To account for this, as has been done by others \citep{ttw02,m02,mkm12}, we use an approximate ``vertically averaged" potential of the form
\begin{equation}
\Phi(s) = G M_p / \sqrt{ s^2 + \delta^2 }
\end{equation}
where we choose $\delta = .6 h$.  The inner and outer radial boundaries are handled via the same damping procedure that was used in \cite{disco}.  As in that work, we do not attempt to model accretion onto the planet (The effects of accretion on the gap formation process have been discussed by \cite{dl08}).

\section{Results}
\label{sec:res}

\subsection{Saturated Gap Depth}

\begin{figure}
\epsscale{1.0}
\plotone{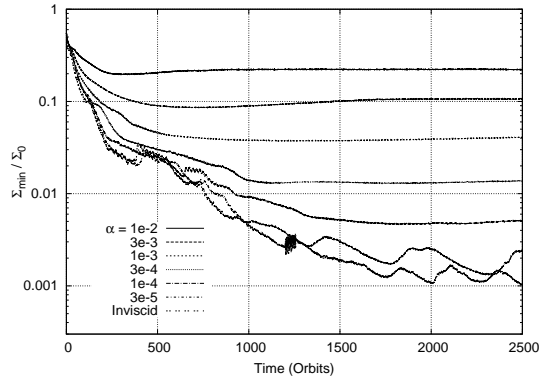}
\caption{ After many orbits, the planet opens a gap, which eventually saturates to its final steady-state depth, at which the rate of viscous filling is equal to the rate of  evacuation.  Plotted here is the (time-averaged) minimum density in the gap, as a function of time, for various disk viscosities.  The final saturated minimum is proportional to the viscosity.  The planet mass in this example is $M_p = 4 M_{Sh}$ (About half of Jupiter's mass).
\label{fig:depthtime} }
\end{figure}

\begin{figure}
\epsscale{1.0}
\plotone{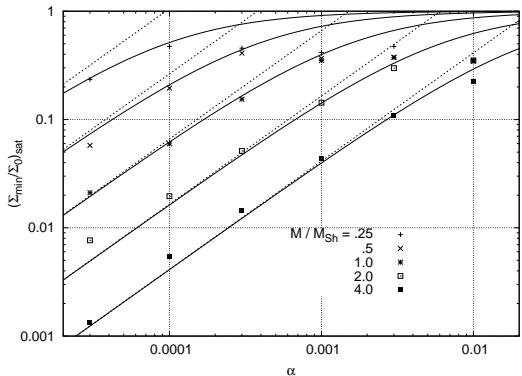}
\caption{ The saturated gap depth obeys simple scaling formulas, as a function of disk viscosity $\alpha$ and the mass of the secondary.  We plot minimum gap density for all test cases.  The dashed curves represent the scaling relation (\ref{eqn:depth}).  Solid curves incorporate the modification (\ref{eqn:mod}).
\label{fig:depth} }
\end{figure}

Each system evolves for thousands of orbits before finally saturating, the secondary having hollowed out a gap with some minimum density.  We plot minimum density on the grid as a function of time for the $4 M_{Sh}$ secondary in Figure \ref{fig:depthtime}, for various viscosities.  For this massive secondary, saturation is established after a few thousand orbits, and the saturated minimum is proportional to the viscosity.  We plot the final $\Sigma$ in all test cases in Figure \ref{fig:depth}.  These can be mapped to simple scaling relations, as a function of ${M_p / M_{Sh}}$ and $\alpha$.  The minimum density is linear in $\alpha$ and inversely quadratic in the secondary mass, as it is a nonlinear effect of the perturbation.  We construct an empirical scaling relation for $\Sigma_{sat}$:

\begin{equation}
\Sigma_{sat} / \Sigma_0 = 29 \mathcal{M} \left({M_p \over M_{Sh}}\right)^{-2} \alpha
\label{eqn:depth}
\end{equation}

We note that the gap depth obeys power-law scaling with respect to viscosity and planet mass.  This result is a bit unexpected from the standpoint of 1D torque-balance models of gap profiles \citep{v04,cm07} which typically predict exponential dependence of gap depth on system parameters.

Equation (\ref{eqn:depth}) should be slightly modified to take into account secondary masses below the threshold, for which this formula would erroneously predict ${\Sigma_{sat} > \Sigma_0}$.  Replacing $\Sigma_{sat}$ with

\begin{equation}
\tilde \Sigma_{sat} = \Sigma_{sat} \Sigma_0 / ( \Sigma_{sat} + \Sigma_0 )
\label{eqn:mod}
\end{equation}

fits the low-mass, high-viscosity data reasonably well (Figure \ref{fig:depth}).  However, the behavior for below-critical perturbers is less important; equation (\ref{eqn:depth}) gives us the machinery to address the question of gap opening.

\subsection{Gap opening criterion}

We choose our diagnostic criterion for gap opening to agree with \cite{cm07}:

\begin{equation}
\Sigma_{sat} < 0.1 \Sigma_0,
\label{eqn:sigma_cond}
\end{equation}

Using (\ref{eqn:depth}) we arrive at the criterion

\begin{equation}
M_{gap} = M_{Sh} ( 17 \sqrt{ \alpha \mathcal{M} } ).
\label{eqn:finalcrit}
\end{equation}

\begin{figure}
\epsscale{1.0}
\plotone{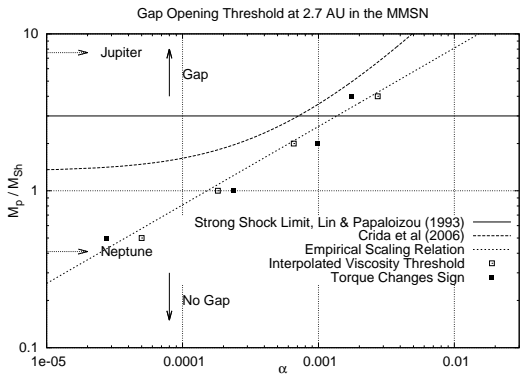}
\caption{ Gap opening criteria, dependent on viscosity and planet mass.  We plot several curves representing potential gap opening criteria.  Planet-Disk systems below a given curve will not open a gap according to the criterion.  We plot equations (\ref{eqn:hillradius}) and (\ref{eqn:crida}), corresponding to predicted gap-opening thresholds, alongside our empirically derived scaling relation (\ref{eqn:finalcrit}).  To highlight the accuracy of our scaling relation, we plot two distinct measured criteria for gap-opening.  Open squares represent the viscosity at which the minimum surface density dips below ${\Sigma < 0.1 \Sigma_0}$, and the filled squares represent the viscosity at which the total torque on the planet changes sign, a noisier measurement which is nevertheless roughly consistent with the minimum surface density measurement.
\label{fig:gapplot} }
\end{figure}

We plot this criterion in Figure \ref{fig:gapplot}, alongside data points which were interpolated from Figure \ref{fig:depth}.  We find excellent agreement.  Alongside this, we plot the ``Strong Shock" prediction of \cite{lp_main}, and the combined prediction of \cite{cm07} (Equation \ref{eqn:crida}).
  We note what appears to be disagreement between these previous predictions and our current empirical results, at least for low viscosities and small planets.  Eq. (\ref{eqn:crida}) asymptotes to a finite value in the low-viscosity limit, corresponding to the ``Strong Shock" limit $M_p > 1.1 M_{Sh}$.  This discrepancy could be numerical in origin; an under-resolved calculation with a small but non-negligible numerical viscosity (${\alpha_{num} \sim 3 \times 10^{-4}}$) would spuriously produce this behavior.  This emphasizes the importance of low numerical viscosity when studying this low-mass planet regime.

The scaling given in (\ref{eqn:finalcrit}) coincides with older estimates for high mass planets \citep{lp86,wh89}.  The idea that this same scaling would apply for low-mass planets was suggested by \cite{r02b}.  It should be noted that in this work we have restricted ourselves to the case ${\mathcal{M}=20}$; the scaling with Mach number should be checked in a future work.  This formula suggests the scaling (\ref{eqn:scale_h}), which implies a fixed gap width ${\Delta \sim h}$.  This can be investigated by examining the gap profile directly.

\subsection{Gap profile}

\begin{figure}
\epsscale{1.0}
\plotone{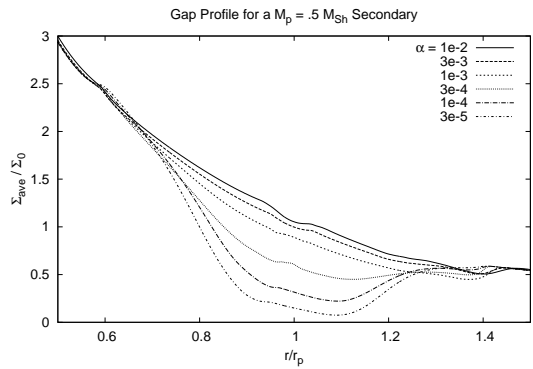}
\caption{ Azimuthally averaged surface density at late times for a planet of roughly Neptune's mass ($M_p = .5 M_{Sh}$).  Curves are plotted with various viscosities.  Though the planet mass is well below the limit of \cite{cm07}, a gap opens at low viscosity (${\alpha \lesssim 10^{-4}}$).
\label{fig:alphprof} }
\end{figure}

We plot the azimuthally averaged density profile for one planet, ${M_p = .5 M_{Sh}}$, with various viscosities, in Figure \ref{fig:alphprof}.  Though it is too small to satisfy the strong shock condition, clearly this secondary opens a gap when the viscosity is low enough.

\begin{figure}
\epsscale{1.0}
\plotone{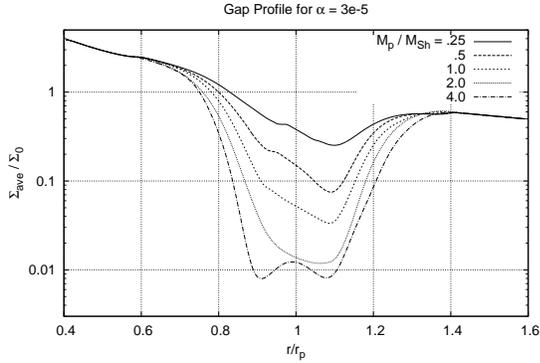}
\caption{ Azimuthally averaged surface density at late times for various planet masses (${\alpha = 3 \times 10^{-5}}$).  For low mass planets, the gap width is weakly dependent on planet mass.
\label{fig:massprof} }
\end{figure}

\begin{figure}
\epsscale{1.0}
\plotone{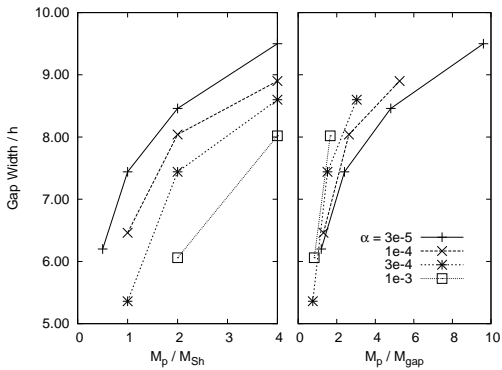}
\caption{ The gap width is given for various planet masses and viscosities.  The edge of the gap is defined as the point where $\Sigma = \Sigma_0 / 3$.  In the right panel, we re-scale the planet mass as a fraction of the critical gap-opening mass (Equation \ref{eqn:finalcrit}).  We find that the curves for various viscosities converge at around $M_p \sim M_{gap}$, with a fixed width of $\Delta \sim 6 h$.
\label{fig:width} }
\end{figure}

The scaling relation (\ref{eqn:finalcrit}), when compared with (\ref{eqn:scale_h}) and (\ref{eqn:scale_r}), seems to imply for low-mass planets that the gap width should be given by the scale height , and not the Hill radius of the secondary.  In order to study this, we plot the gap profile in the low-viscosity limit for our various planet masses (Figure \ref{fig:massprof}).  Though the largest mass is 16 times the smallest mass, there is not a significant difference in gap width between the different profiles.  However, as there is some small variation in gap width, we plot its dependence on planet mass in Figure \ref{fig:width}.  Here, the edge of the gap is defined as the radial position at which $\Sigma < \Sigma_0/3$.  The width is plotted in Figure \ref{fig:width} for various masses and viscosities, and it is clear there is a weak dependence on planet mass.  In the right panel of this figure, each curve is re-scaled so that the planet mass is expressed as a fraction of $M_{gap}$ (Equation \ref{eqn:finalcrit}).  Interestingly, the curves appear to converge at the critical mass $M_p = M_{gap}$, at a fixed width of $\Delta \sim 6 h$.  This suggests that the shape of the gap profile at the critical mass may be roughly independent of other parameters of the system.  \cite{r02} suggests that the gap edge should coincide with wherever the density wave dissipates (where it shocks, in the inviscid case).

\subsection{Stability of Deep Gaps}
\label{sec:vortex}

Looking again at Figure \ref{fig:depthtime}, we note that the inviscid calculation is plotted alongside the viscous results, apparently having a slightly lower effective viscosity than the ${\alpha = 3 \times 10^{-5}}$ run.  Naively, we might be tempted to interpret this effective viscosity as numerical, but we shall see in the next section that our numerical viscosity is ${\alpha_{num} \sim 10^{-6}}$, which is much too small for this interpretation.

A hint as to the explanation can be seen in Figure \ref{fig:depthtime} at around 1200 orbits.  The inviscid calculation has some intermittent, erratic behavior which settles down on roughly a 10 orbit timescale.  This short-timescale jostling occurs about every thousand orbits or so, and it is due to vortices forming and breaking up after several orbits.  These vortices seem to provide an effective viscosity which appears to be the cause of the saturation floor; When we double the resolution this saturation floor does not change.  Similar vortices have been witnessed and analyzed in previous planet-disk calculations \citep{ll09,yl10}.  In particular, these works demonstrated the impact such vortices can have on planet migration.

\subsection{Resolution Study}

\begin{figure}
\epsscale{1.0}
\plotone{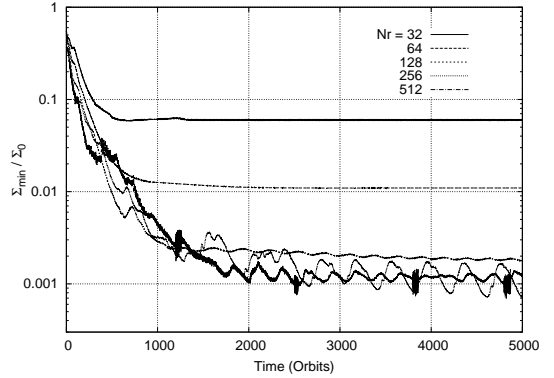}
\caption{ Inviscid calculations, showing dependence of gap depth on resolution ($M_p = 4 M_{Sh}$).  For ${N_R = 32}$ and $64$, numerical viscosity dominates.  For higher resolutions, a saturated gap depth is shown which is not caused by explicit viscosity or numerical viscosity.  It appears to be caused by a fluid instability which generates vortices that drift and dissipate, preventing further gap deepening.
\label{fig:resstudy} }
\end{figure}

To determine our effective numerical viscosity for this problem, we performed several inviscid tests for the $M_p = 4 M_{Sh}$ secondary at lower resolution, comparing 32, 64, 128, and 256 radial zones to our fiducial 512 radial zone calculation.  The time history of gap depth for these resolutions is plotted in Figure \ref{fig:resstudy}.  We compare this with Figure \ref{fig:depthtime}.  For the cases ${N_R = 32}$ and $64$, the saturated gap depth is clearly a result of numerical viscosity.  However, for higher resolutions, the solution converges on a gap depth.  We discussed this convergence in \S \ref{sec:vortex}; it appears to be the result of unstable vortices providing an effective viscosity.  In Figure \ref{fig:resstudy} the 512 calculation clearly has intermittent oscillations, indicating that this instability appears to re-activate every thousand orbits or so.

Results from the ${N_R = 32, 64}$ and $128$ runs can be used to construct a formula for saturated density as a function of the resolution.  This yields a formula for effective numerical viscosity by comparing our result with Eq. \ref{eqn:depth}.  The effective viscosity we find is
\begin{equation}
\alpha_{num} = 2.5 \times 10^{-3} \left({\Delta r \over h }\right)^2
\end{equation}
where $\Delta r$ measures the radial size of a zone near the orbital radius.

Our numerical viscosity for the ${N_R = 512}$ runs is therefore approximately ${\alpha_{512} \sim 4 \times 10^{-6}}$, roughly an order of magnitude below the physical viscosity we introduce, and also smaller than the effective viscosity produced by vortices through shear instabilities in the inviscid case.  We should note, of course, that this measure of numerical viscosity is not only problem-dependent, but diagnostic-dependent.  That is, it is expected that some regions of the computational domain have more numerical viscosity than others, partly because of the variable resolution but partly also because the numerical viscosity is probably larger near shocks (the scaling of numerical viscosity would also be linear near shocks).  That caveat aside, this diagnostic is a reasonable measure of ``effective viscosity for this problem".

\section{Summary}
\label{sec:sum}

We have performed 2D numerical calculations which directly demonstrate that the ``strong shock limit" (\ref{eqn:stab}) proposed by \cite{lp_main}, and used by \cite{cm07} is not a necessary condition for gap-opening.  While our results show reasonable agreement with \cite{cm07} for large viscosity and planet mass, we do not agree for low viscosities and masses.  In particular, Neptune-mass planets are capable of gap opening for $\alpha \lesssim 10^{-4}$, which is not allowed according to these previously proposed criteria.  We also find that the critical mass scales as the square root of viscosity when applied to low-mass planets.  Generally, we do not find any minimum mass limit for the case $\alpha \rightarrow 0$.

Because gap opening is still possible when the Hill radius is smaller than a scale height, we expect that 3D calculations will be necessary to capture disk-planet dynamics in this low-mass regime.  Since we have found no ``strong shock" limit for gap opening, this means that the inertial limit (Equation \ref{eqn:inertial}) is important in low-viscosity disks.  Up to now, we have considered the case where $\Sigma$ was negligible in the sense that the torque on the planet did not change its orbit; we kept both primary and secondary on fixed circular orbits about the center of mass.   In order to determine the inertial limit in the low-mass regime, we must relax this assumption and allow the planetary orbit to respond to the torque exerted by the disk.  All of this is planned as future work. 

\acknowledgments
This research was supported in part by NASA through grants NNX10AF62G and NNX11AE05G issued through the Astrophysics Theory Program and by the NSF through grant AST-1009863.  Resources supporting this work were provided by the NASA High-End Computing (HEC) Program through the NASA Advanced Supercomputing (NAS) Division at Ames Research Center.  We are grateful to Roman Rafikov and Andrei Gruzinov for many helpful comments and discussions.  We also thank the anonymous referee for his/her thorough review.

\begin{appendix}
\section{Viscosity Implementation in DISCO}
\label{app}

The viscous terms in DISCO are calculated as viscous fluxes, which requires calculating the divergence of the viscous stress tensor in cylindrical coordinates.  DISCO automatically takes into account the volume element in cylindrical coordinates, because we use exact geometry to calculate the volumes of cells and areas of faces.  Therefore, we merely need to re-write the viscous stresses $T_{ij}$ in terms of the ``effective stress tensor" $T^{eff}$, defined by:

\begin{equation}
(\nabla \cdot T)_j = {1 \over \sqrt{g}}\partial_r ( \sqrt{g} T^{eff}_{r j} ) + {1 \over \sqrt{g}} \nabla_{\phi} ( \sqrt{g} T^{eff}_{\phi j} ) + S_j
\end{equation}

The components of $T^{eff}$ are the viscous fluxes as they appear in the code.  Note that when we express the divergence of the stress tensor in cylindrical coordinates, we generate a source term in the radial component of momentum, for the same reason that a centrifugal source term is generated when taking the divergence of the hydrodynamical stress tensor.  This source term will have a negligible effect on disk dynamics, but is taken into account in the code for completeness.  The components of $T^{eff}$ are

\begin{equation}
\begin{array}{rcl}
T^{eff}_{r r} & = & -(\rho \nu) \nabla_r v_r \\
T^{eff}_{r \phi} & = & -(\rho \nu) ( \nabla_{\phi} v_r - 2 \Omega ) \\
T^{eff}_{\phi r} & = & -(\rho \nu) ( r^2 \nabla_{r} \Omega ) \\
T^{eff}_{\phi \phi} & = & -(\rho \nu) ( r^2 \nabla_{\phi} \Omega + 2 v_r )
\label{eqn:teff}
\end{array}
\end{equation}

Where $v_r$ is the radial velocity and $\Omega$ is the angular frequency.  The first two components listed are the flux of the radial component of momentum, and the remaining two components are flux of angular momentum.  The source term for the momentum in the radial direction is

\begin{equation}
S_r = -(\rho \nu) v_r/r^2.
\label{eqn:tsrc}
\end{equation}

Following a similar path to \cite{mvisc}, we calculate the viscous flux independently from the hydrodynamic flux (the hydro flux uses a Riemann solver).  In order to evaluate (\ref{eqn:teff}) we need to know the gradients of primitive variables.  For this, we use the same slope-limited gradients which are used to extrapolate primitive variables from zone centers to faces.  We extrapolate primitive variables to either side of each face, as is done for the hydro fluxes.  We then evaluate the arithmetic mean of the variables on either side of the face, and of the gradients, and use these averaged primitive variables and gradients to calculate the fluxes given in (\ref{eqn:teff}).  The source term (\ref{eqn:tsrc}) is simply calculated as a cell-centered quantity.

In contrast with \cite{mvisc}, we do not need to extrapolate our variables in time for a half-timestep, because our time integration is based on the method of lines; we achieve high order by performing the full time integration via a series of Runge-Kutta timesteps.  Each individual step can therefore be first-order in time.  This simplification means that our viscosity implementation can be simply described by the equations (\ref{eqn:teff}) and (\ref{eqn:tsrc}).  In principle, we should have to take a shorter timestep if the characteristic viscous timescale for a zone is shorter than its sound-crossing time, but in this work we deal with viscosities small enough that we do not need to consider this timescale.

\end{appendix}

{}

\end{document}